\documentclass[11pt]{amsart}

\usepackage{amssymb}
\usepackage{epsf}
\usepackage{a4wide}

\usepackage[pdftitle={Kolakoski-(2m,2n) are limit-periodic model
 sets},pdfauthor={Bernd Sing},pdfsubject={Mathematical
 Physics},dvipdfm]{hyperref} 

\newtheorem{theorem}{Theorem}
\newtheorem{lem}{Lemma}
\newtheorem{prop}{Proposition}

\newtheoremstyle{citing}
  {3pt}
  {3pt}
  {\itshape}
  {}
  {\bfseries}
  {.}
  {.5em}
  {\thmnote{#3}}

\theoremstyle{citing}
\newtheorem*{varthm}{}

\newcommand{\NN}{\mathbb{N}}
\newcommand{\ZZ}{\mathbb{Z}}
\newcommand{\QQ}{\mathbb{Q}}
\newcommand{\RR}{\mathbb{R}}
\newcommand{\CC}{\mathbb{C}}
\newcommand{\bs}[1]{\boldsymbol{#1}}
\newcommand{\ABC}{\mathcal{A}}
\newcommand{\HS}{\mathcal{L}_{}^{2}(X(\varrho),\mu)}
\newcommand{\DS}[1]{(X(#1),T)}

\begin{document}

\title{Kolakoski-$(2m,2n)$ are limit-periodic model sets}  

\author{Bernd Sing}
\address{Institut f\"{u}r Mathematik, Universit\"{a}t Greifswald,
Jahnstr.~15a, 17487 Greifswald, Germany}
\email{\htmladdnormallink{sing@uni-greifswald.de}{mailto: sing@uni-greifswald.de}}
\urladdr{\htmladdnormallink{http://schubert.math-inf.uni-greifswald.de}{http://schubert.math-inf.uni-greifswald.de}} 

\begin{abstract} 
We consider (generalized) Kolakoski sequences on an alphabet with two even
numbers. They can be related to a primitive substitution rule of constant
length $\ell$. Using this connection, we prove that they have pure point
dynamical and pure point diffractive spectrum, where we make use of the 
strong interplay between these two concepts. Since these sequences can then be
described as model sets with $\ell$-adic internal space, we add an approach to
``visualize'' such internal spaces.
\end{abstract}
\maketitle

\section{Introduction}
 
A one-sided infinite sequence $\omega$ over the alphabet $\ABC=\{1,2\}$ is
called a (classical) \textit{Kolakoski sequence} (named after W.~Kolakoski who
introduced it in 1965, see~\cite{Kol65}), if it equals the sequence defined by
its run lengths, e.g.:
\begin{equation}\label{eq:kol}
\begin{array}{ccccccccccccccc}
\omega & = & \underbrace{22} & \underbrace{11} & \underbrace{2} &
\underbrace{1} & \underbrace{22} & \underbrace{1} & \underbrace{22} &
\underbrace{11} & \underbrace{2} & \underbrace{11} & \ldots && \\
&& 2 & 2 & 1 & 1 & 2 & 1 & 2 & 2 & 1 & 2 & \ldots & = & \omega.
\end{array}
\end{equation}
Here, a \textit{run} is a maximal subword consisting of identical letters. The
sequence $\omega'=1\omega$ is the only other sequence which has this property.

One way to obtain $\omega$ of (\ref{eq:kol}) is by starting with $2$ as a seed
and iterating the two substitutions 
\begin{equation*}
\sigma^{}_0: \begin{array}{lcl} 1 & \mapsto & 2 \\ 2 & \mapsto & 22 \end{array}
\quad \text{and} \quad
\sigma^{}_1: \begin{array}{lcl} 1 & \mapsto & 1 \\ 2 & \mapsto & 11,
\end{array} 
\end{equation*}
alternatingly, i.e., $\sigma^{}_0$ substitutes letters on even positions and
$\sigma^{}_1$ letters on odd positions (we begin counting at $0$):
\begin{equation*}
2 \mapsto 22 \mapsto 2211 \mapsto 221121 \mapsto 221121221 \mapsto \ldots
\end{equation*}
Clearly, the iterates converge to the Kolakoski sequence $\omega$ (in the
obvious product topology), and $\omega$ is the unique (one-sided) fixed point
of this iteration.  

One can generalize this by choosing a different alphabet
$\ABC=\{p,q\}$ (we are only looking at alphabets with $\operatorname{card}
(\ABC) = 2$). Such a (generalized) Kolakoski sequence, which is also equal to
the sequence of its run lengths, can be obtained by iterating the two
substitutions 
\begin{equation}\label{eq:subs0}
\sigma^{}_0: \begin{array}{lcl} q & \mapsto & p^q \\ p & \mapsto & p^p
\end{array} \quad \text{and} \quad
\sigma^{}_1: \begin{array}{lcl} q & \mapsto & q^q \\ p & \mapsto & q^p
\end{array} 
\end{equation}
alternatingly. Here, the starting letter of the sequence is $p$. We will call
such a sequence a Kolakoski-$(p,q)$ sequence, or Kol$(p,q)$ for short. The
classical Kolakoski sequence $\omega$ of (\ref{eq:kol}) is therefore denoted
by Kol$(2,1)$ (and $\omega'$ by Kol$(1,2)$).  

While little is known about the classical Kolakoski sequence
(see~\cite{Dek97}), and the same holds for all Kol$(p,q)$ with $p$ odd and $q$
even or vice versa (see~\cite{Diplom}), the situation is more favorable if
$p$ and $q$ are either both even or both odd. If both are odd, one can, in some
cases, rewrite the substitution as a substitution of Pisot type
(see~\cite{Diplom, BS02}), which can be described as (limit--) aperiodic model
sets. By this method, Kol$(3,1)$ is studied in~\cite{BS02} and shown to be a
deformed model set. The case where both symbols are even will be studied below.

It is the aim of this article to determine structure and order
of the sequences Kol$(2m,2n)$. This will require two steps: First we establish
an equivalent substitution of constant length for Kol$(2m,2n)$ and analyze it
with methods known from the theory of dynamical systems. Then we conclude
diffractive properties from this.

\medskip
\noindent
\textsc{Remark}: Every Kol$(p,q)$ can uniquely be extended to a
bi-infinite (or two-sided) sequence. The one-sided sequence (to the right) is
Kol$(p,q)$ as explained above. The added part to the left is a reversed copy
of Kol$(q,p)$, e.g., in the case of the classical Kolakoski sequence of
(\ref{eq:kol}), this reads as
\begin{equation*}
\ldots 11221221211221|22112122122112 \ldots,
\end{equation*}
where ``$|$'' denotes the seamline between the one-sided sequences. Note that,
if $q=1$ (or $p=1$), the bi-infinite sequence is mirror symmetric around the
first position to the left (right) of the seamline. The bi-infinite sequence
equals the sequence of its run lengths, if counting is begun at the seamline. 
Alternatively, one can get such a bi-infinite sequence by starting with $q|p$
and applying the two substitutions to get $\sigma^{}_1(q)|\sigma^{}_0(p)$ in
the first step and so forth. This also implies that Kol$(p,q)$ and Kol$(q,p)$
will have the same spectral properties, and it suffices to study one of them.

\bigskip

\section{Kol\texorpdfstring{$(2m,2n)$}{(2m,2n)} as Substitution of Constant
  Length} 

If both letters are even numbers, i.e., $p=2m$ and $q=2n$ (with $m\neq n$,
where we can concentrate on $m > n$ by the above discussion),
one can build blocks of two letters and  obtain an (ordinary)
substitution. Setting $A= pp$ and $B=qq$, these substitutions and their
\textit{substitution matrix} $\bs{M}$ (sometimes called \textit{incidence
  matrix} of the substitution) are given by 
\begin{equation} \label{eq:subs1}
\sigma: \begin{array}{lcl} A & \mapsto & A_{}^{m} B_{}^{m} \\ B & \mapsto &
  A_{}^{n} B_{}^{n} \end{array} \quad \text{and} \quad \bs{M}=\left(
  \begin{array}{cc} m & m \\ n & n \end{array} \right),
\end{equation}
where the entry $M_{ij}$ is the number of occurrences of $j$ in $\sigma(i)$
($i,j \in \{A,B\}$; sometimes the transposed matrix is used). A bi-infinite
fixed point can be obtained as follows:
\begin{equation*}
B|A \mapsto A^nB^n|A^mB^m \mapsto \ldots
\end{equation*}
This corresponds to the unique bi-infinite Kol$(2m,2n)$ according to our above
convention. 

A substitution $\varrho$ is \textit{primitive} if the corresponding
substitution matrix $\bs{M}$ is primitive, i.e., $\bs{M}^{k}$ has positive
entries only for some $k \in \NN$. Equivalently, $\varrho$ is primitive if
there exists a positive integer $k \in \NN$ such that every $i \in \ABC$
occurs in $\varrho_{}^{k}(j)$ for all $j \in \ABC$. The vector $\bs{\ell}$ with
components $\ell^{}_{i} = |\varrho(i)|$, for $i \in \ABC$, is called the
\textit{length} of the substitution $\varrho$. If all $\ell^{}_{i}$ are equal,
$\varrho$ is a substitution of \textit{constant length}. For the substitution
$\sigma$ of~(\ref{eq:subs1}), we have 
\begin{equation*}
\bs{\ell} = \left( \begin{array}{c} 2m \\ 2n \end{array} \right) ,
\end{equation*}
which is therefore not of constant length (recall that $m \neq n$). 

We will also need some notions from the theory of \textit{dynamical systems},
see~\cite{Que87} and \linebreak 
\cite[Chapters 1, 5 and 7]{BFMS02} for details. Let
$\varrho$ be a primitive substitution over $\ABC$ and $u \in \ABC_{}^{\ZZ}$ a
bi-infinite fixed point of $\varrho$ (i.e., $u=\varrho_{}^{k}(u)$ for some $k
\in \NN$). Denote by $u^{}_{k}$ the $k$th letter of $u$ ($k \in \ZZ$) and by
$T$ the \text{shift map} (i.e., $(T(u))^{}_{k} = u^{}_{k+1}$). Let $\ABC$ be
equipped with the discrete and $\ABC_{}^{\ZZ}$ with the corresponding product
topology. If we set
\begin{equation*}
X(\varrho) = \overline{\{T_{}^{k}(u) \mathbin| k \in \ZZ \}},
\end{equation*}
then $\DS{\varrho}$ is a \textit{dynamical system}. Since we require
$\varrho$ to be primitive, this dynamical system is \textit{minimal} (i.e.,
$\{T_{}^{k}(u)\mathbin| k \in \ZZ\}$ is dense in $X(\varrho)$ for all $u \in
\DS{\varrho}$), does not depend on the chosen fixed point $u$ (if more than
one exists, which is possible in the two-sided situation) and has a unique
probability measure $\mu$ associated with it. In other words, it is
\textit{strictly ergodic}. On the Hilbert space $\HS$, we
have the unitary operator 
\begin{equation*}
\begin{array}{llcl}
U: & \HS & \to & \HS, \\
& f & \mapsto & f \circ T.
\end{array}
\end{equation*}
If $Uf=e_{}^{i \lambda}f$ for some $0 \neq f \in \HS$, we call
$e_{}^{i \lambda}$ an \textit{eigenvalue} of $\DS{\varrho}$ and $f$ the
corresponding \textit{eigenfunction}. The \textit{spectrum} (of the dynamical
system) is said to be a \textit{pure point dynamical spectrum} (or
\textit{discrete spectrum}),  if the eigenfunctions span $\HS$. If $1$ is the 
only eigenvalue and the only eigenfunctions are the constants, the spectrum is 
\textit{continuous}. It is also possible that it has pure point and continuous
components. In that case it is called \textit{partially continuous}. Two
dynamical systems $(X, T)$ and $(Y, S)$ are \textit{isomorphic} (or
\textit{measure-theoretically isomorphic}), if there
exists an invertible measurable map $\varphi: X \to Y$, almost everywhere
defined, such that $\varphi$ preserves the measure and the dynamics (i.e.,
$\varphi \circ T = S \circ \varphi$). 

The spectral theory of primitive substitutions of constant length is well
understood. By the following criterion, we know that the substitutions $\sigma$
of~(\ref{eq:subs1}) are related to substitutions of constant length. 

\begin{lem}\label{lem:Dek-V1}\cite[Section V, Theorem 1]{Dek78}
Let $\varrho$ be a substitution of nonconstant length $\bs{\ell}$. If
$\bs{\ell}$ is a right eigenvector of the corresponding substitution matrix
$\bs{M}$, then $(X(\varrho),T)$ is isomorphic to a substitution dynamical
system generated by a substitution of constant length. \qed
\end{lem}   

Since the substitutions $\sigma$ of~(\ref{eq:subs1})
  fulfill\footnote{\label{foo:note}Note 
  that from~(\ref{eq:subs0}), one can also construct a primitive substitution
  by distinguishing odd and even positions, e.g., for Kol$(4,2)$ we would get
  (we use $\tilde{\cdot}$ as mark for even positions)
\begin{equation*}
\begin{array}{lcl}
4 & \to & 4\tilde{4}4\tilde{4} \\
\tilde{4} & \to & 2\tilde{2}2\tilde{2} \\
2 & \to & 4\tilde{4} \\
\tilde{2} & \to & 2\tilde{2}. 
\end{array}
\end{equation*}
Instead of~(\ref{eq:subs1}), we would get a substitution with substitution
matrix 
\begin{equation*}
\bs{M} = \left(
\begin{array}{cccc}
m & m & 0 & 0 \\
0 & 0 & m & m \\
n & n & 0 & 0 \\
0 & 0 & n & n
\end{array}\right),
\end{equation*}
which is also primitive ($\bs{M}^2$ has positive entries only), but does not
fulfill the requirements of Lemma~\ref{lem:Dek-V1}. The eigenvalues of this
$\bs{M}$ are $\{0,0,0,m+n\}$.} the requirements of this lemma, the next
\linebreak  
task is now to construct the corresponding substitutions of constant
length. This is \linebreak 
achieved by numbering the $A$'s and
$B$'s in~(\ref{eq:subs1}), i.e., we make the substitutions $A_{}^{m} B_{}^{m}
\to A^{}_1 \ldots A^{}_m B^{}_1 \ldots B^{}_m$, respectively 
$A_{}^{n} B_{}^{n} \to A^{}_{m+1} \ldots A^{}_{m+n} B^{}_{m+1} \ldots
B^{}_{m+n}$. Then, the former substitutions~(\ref{eq:subs1}) induce  
\begin{equation}\label{eq:subs2}
\begin{array}{lcl}
\scriptstyle{A^{}_1 \ldots A^{}_m B^{}_1 \ldots B^{}_m} &
\scriptstyle{\mapsto} & \scriptstyle{(A^{}_1 \ldots A^{}_m B^{}_1 \ldots
  B^{}_m)_{}^{m} (A^{}_{m+1} \ldots A^{}_{m+n} B^{}_{m+1} \ldots
  B^{}_{m+n})_{}^{m}} \\ 
\scriptstyle{A^{}_{m+1} \ldots A^{}_{m+n} B^{}_{m+1} \ldots B^{}_{m+n}} &
\scriptstyle{\mapsto} & \scriptstyle{(A^{}_1 \ldots A^{}_m B^{}_1 \ldots
  B^{}_m)_{}^{n} (A^{}_{m+1} \ldots A^{}_{m+n} B^{}_{m+1} \ldots
  B^{}_{m+n})_{}^{n}.} 
\end{array}
\end{equation}
From this we get substitutions of constant length $m+n$ (the eigenvalue of the
substitution matrix $\bs{M}$ in~(\ref{eq:subs1})) by parting the right sides
in blocks of $m+n$ letters. For example, let $m=2$ and $n=1$. Then
\begin{equation*}
\begin{array}{lcl}
A^{}_1 A^{}_2 B^{}_1 B^{}_2 & \mapsto & A^{}_1 A^{}_2 B^{}_1 \quad B^{}_2
A^{}_1 A^{}_2 \quad B^{}_1 B^{}_2 A^{}_3 \quad B^{}_3 A^{}_3 B^{}_3 \\
A^{}_3 B^{}_3 & \mapsto & A^{}_1 A^{}_2 B^{}_1 \quad B^{}_2 A^{}_3 B^{}_3 
\end{array}
\end{equation*}
and one extracts the following substitution of constant length $3$:
\begin{equation}\label{eq:kol42h2}
\begin{array}{lcl}
A^{}_1 & \mapsto & A^{}_1 A^{}_2 B^{}_1 \\
A^{}_2 & \mapsto & B^{}_2 A^{}_1 A^{}_2 \\
B^{}_1 & \mapsto & B^{}_1 B^{}_2 A^{}_3 \\
B^{}_2 & \mapsto & B^{}_3 A^{}_3 B^{}_3 \\
A^{}_3 & \mapsto & A^{}_1 A^{}_2 B^{}_1 \\ 
B^{}_3 & \mapsto & B^{}_2 A^{}_3 B^{}_3. 
\end{array} 
\end{equation}
In the same way, we get substitutions of constant length $m+n$
from~(\ref{eq:subs2}). Note that these substitutions are all primitive
since (compare to~(\ref{eq:subs2})) in every block of $2m^2+2n$ successive
letters (note that we use $m>n$) every letter of
$\ABC=\{A^{}_1,\ldots,A^{}_{m+n},B^{}_1,\ldots,B^{}_{m+n}\}$ occurs, so if
$(m+n)^{k_0} \ge 2m^2+2n$, then $\bs{M}^{k_0}$ has positive entries only (this
holds for $k_0 \ge3$). Note also that we can reduce the alphabet by one
letter by identifying $A^{}_1 = A^{}_{m+1}$ ($A^{}_1 = A^{}_3$ in the
example~(\ref{eq:kol42h2})), because both $A$'s always yield the same
substitution. 
 
Let us now determine the positions of $A^{}_1$ in the sequence $u$ generated
by~(\ref{eq:subs2}). They are given by $\alpha \cdot 2m+\beta \cdot 2n$ for
some $\alpha,\beta \in \ZZ$ (e.g., $0,\, 2m,\, 4m,\ldots, 2m^2,\, 2m^2+2n,\,
2m^2+4n, \ldots 2m^2+2nm, \ldots$). Therefore we get $\gcd\{i\mathbin|
u^{}_{i} = u^{}_{0} = A^{}_1 \} = \gcd(2m,2n) = 2 \gcd(m,n)$. The
\textit{height} $h(\varrho)$ of a primitive substitution $\varrho$ of constant
length $\ell$ which generates a sequence $u$ is defined as
\begin{equation}\label{eq:def_height}
h(\varrho) = \max\{k\ge1\mathbin| \gcd(k,\ell)=1 \text{ and } k \text{ divides
    } \gcd\{i\mathbin|u^{}_{i}=u^{}_0\}\}.
\end{equation}
Then the following lemma holds.

\begin{lem}\label{lem:Dek-II13}
Let $\DS{\varrho}$ be a dynamical system, where $\varrho$ is a primitive
substitution of constant length $\ell$ and height $h(\varrho)$. Then the
pure point part of this dynamical system is isomorphic to the dynamical system
$(\ZZ^{}_{\ell} \,\times\, \ZZ/h(\varrho)\ZZ, \tau)$, where $\tau$ is the
addition of $(1,1)$ on the Abelian group $\ZZ^{}_{\ell} \,\times\,
\ZZ/h(\varrho)\ZZ$, i.e., the direct product of the $\ell$-adic integers
$\ZZ^{}_{\ell}$ and the cyclic group $\ZZ / h(\varrho)\ZZ$ of order
$h(\varrho)$. Therefore the pure point dynamical spectrum is given by 
\begin{equation*}
\left\{ \left. e^{2 \pi i \frac{n}{\ell^m}+2\pi i
  \frac{k}{h(\varrho)}}\,\right|\, k,n \in \ZZ, m \in \NN \right\}.
\end{equation*}
\end{lem}
\noindent
Note that 
\begin{equation}\label{eq:ladic}
\ZZ^{}_{\ell} \simeq \ZZ^{}_{p^{}_1} \times \ldots \times
\ZZ^{}_{p^{}_{r}}, 
\end{equation}
where $p^{}_1, \ldots, p^{}_{r}$ are the distinct primes dividing $\ell$,
see~\cite[Section 3.10]{Mar71}. 
\begin{proof}
The lemma is just a reformulation of~\cite[Theorem II.13]{Dek78}, compare
with~\cite[Section VI.]{Que87} and~\cite[Section 7.3]{BFMS02}.
\end{proof}

\begin{prop}\label{prop:spectrum}
Suppose $\DS{\sigma}$ has pure point dynamical spectrum, where $\sigma$ is the
substitution of~\textnormal{(\ref{eq:subs1})}. Then 
\begin{equation*}
\DS{\sigma} \simeq \left\{ 
\begin{array}{ll} 
(\ZZ^{}_{m+n}\,\times\,\ZZ/2\ZZ,\tau) & \text{if }m+n\text{ is odd}  \\
(\ZZ^{}_{m+n}, \tilde{\tau}) & \text{if }m+n\text{ is even},
\end{array} \right.
\end{equation*}
where $\tau$ is the addition of $(1,1)$ and $\tilde{\tau}$ the addition of $1$.
\end{prop}

\begin{proof}
For the substitution~(\ref{eq:subs2}) of constant length $\ell = m+n$, we have
already seen that $\gcd\{i\mathbin| u^{}_{i}=u^{}_0\} = 2\gcd(m,n)$. Therefore,
using~(\ref{eq:def_height}), the height of this substitution is $2$ if $\ell$
is odd and $1$ if $\ell$ is even. The dynamical system of the
substitution~(\ref{eq:subs2}) is isomorphic to $\DS{\sigma}$ by
Lemma~\ref{lem:Dek-V1}, therefore they have the same spectrum. The remaining
statement follows from Lemma~\ref{lem:Dek-II13}.
\end{proof} 

We want to show that the spectrum of $\sigma$ is indeed pure point. For
this we use slightly different substitutions of constant length that we
deduce from $\sigma$. We substitute $A_{}^{m}B_{}^{m} \to a^{}_1 \ldots
a^{}_{m}$ and $A_{}^{n}B_{}^{n} \to b^{}_1 \ldots b^{}_{n}$ (so
in~(\ref{eq:subs2}) we build essentially blocks of two, e.g., $a^{}_1 =
A^{}_1A^{}_2$). We get
\begin{equation}\label{eq:subs3}
\begin{array}{lcl}
a^{}_1 \ldots a^{}_m & \mapsto & (a^{}_1 \ldots a^{}_m)_{}^{m} (b^{}_1
\ldots b^{}_{n})_{}^{m} \\ 
b^{}_1 \ldots b^{}_{n}  & \mapsto & 
(a^{}_1 \ldots a^{}_m)_{}^{n} (b^{}_1 \ldots b^{}_{n})_{}^{n},
\end{array}
\end{equation}
which again gives substitutions of constant length $\ell = m+n$. They are all
primitive substitutions by the same argument as before (in every block of
$m^2+n$ successive letters every letter occurs). In the case $n>1$, we can
reduce the alphabet by one letter by identifying $a^{}_1 = b^{}_1 \mapsto
a^{}_1 \ldots a^{}_{m} a^{}_1 \ldots a^{}_{n}$. So we have two cases, $n=1$
with substitutions
\begin{equation}\label{eq:kol2m2}
\tilde{\theta}:\left\{ \begin{array}{lclllllll}
a^{}_1 & \mapsto & a^{}_1 & a^{}_2 & a^{}_3 & \ldots & a^{}_{m-1} & a^{}_{m} &
a^{}_1 \\
a^{}_2 & \mapsto & a^{}_2 & a^{}_3 & a^{}_4 & \ldots & a^{}_{m} & a^{}_1 &
a^{}_2 \\
\vdots & & \ldots &&&&&& \\
a^{}_{m-1} & \mapsto & a^{}_{m-1} & a^{}_{m} & a^{}_1 & \ldots & a^{}_{m-3} &
a^{}_{m-2} & a^{}_{m-1} \\
a^{}_{m} & \mapsto & a^{}_{m} & b^{}_1 & b^{}_1 & \ldots & b^{}_1 & b^{}_1 &
b^{}_1 \\ 
b^{}_1 & \mapsto & a^{}_1 & a^{}_2 & a^{}_3 & \ldots & a^{}_{m-1} & a^{}_{m} &
b^{}_1 \\ 
\end{array} \right.
\end{equation}
and $n>1$ with substitutions $\theta$ (it is cumbersome to write down such a
$\theta$ in general form, but we will investigate its structure in the next
section). Now the height of $\theta$ and $\tilde{\theta}$ is always $1$,
because if $n>1$ we get $\gcd\{i\mathbin|u^{}_0=u^{}_{i}=a^{}_1\} = \gcd(m,n)$,
and if $n=1$ we get $\gcd\{i\mathbin|u^{}_0=u^{}_{i}=a^{}_1\} = \gcd(m,m+1) =
1$. 

Let $\varrho$ be a primitive substitution of constant length $\ell$ and height
$h(\varrho)=1$. One says that $\varrho$ admits a \textit{coincidence}, if there
exist a $k \in \NN$ and $j<\ell_{}^{k}$ such that $\varrho_{}^{k}(i)^{}_{j}$
is the same for all $i \in \ABC$ (the $j$th letter of each
$\varrho_{}^{k}(i)$ is the same, i.e., $\varrho_{}^{k}$ admits a column of
identical values). 

\begin{lem}\label{lem:DekIII7}\cite[Section III, Theorem 7]{Dek78}
Let $\DS{\varrho}$ be a substitution dynamical system of constant length and
height $h(\varrho)=1$. Then $\DS{\varrho}$ has pure point dynamical spectrum
if and only if $\varrho$ admits a coincidence.\qed
\end{lem}

If a substitution has height $h>1$, one gets a substitution of
height $1$ by combining letters into blocks of $h$ letters. If this new
substitution has pure point dynamical spectrum, so has the original
substitution of height $h$, see~\cite{Dek78}. Obviously, we get the following:
if the substitutions $\theta$ and $\tilde{\theta}$ (which arise
from~(\ref{eq:subs3})) admit coincidences, then the dynamical systems defined
by $\sigma$ of~(\ref{eq:subs1}) have pure point dynamical spectrum. 

\bigskip

\section{Coincidences and Coincidence Matrix}

Let us first check $\tilde{\theta}$ of~(\ref{eq:kol2m2}) for
coincidences. For this we begin by exploring the structure:
$\tilde{\theta}(a^{}_1)$ has two $a^{}_1$'s at position $0$ and $m$,
$\tilde{\theta}(a^{}_2)$ has an $a^{}_1$ at position $m-1$, etc. We get an
$a^{}_1$ in $\tilde{\theta}(a^{}_{k})$ at position $m+1-k$ for $1\le k \le
m-1$. Similar arguments show that there is an $a^{}_{m}$ in
$\tilde{\theta}(a^{}_{k})$ at position $m-k$ for $1 \le k \le m$ and at $m-1$
in $\tilde{\theta}(b^{}_1)$. Now, 
$\tilde{\theta}(a^{}_{m})$ has $b^{}_1$'s at all positions $1, 
\ldots,m$. Furthermore, $\tilde{\theta}(b^{}_1)$ has a $b^{}_1$ at position $m$
and shares the first $m$ letters with $\tilde{\theta}(a^{}_1)$. Schematically,
we get the following structure of $\tilde{\theta}$:
\begin{equation}\label{eq:kol2m2_e}
\begin{array}{lcccccccc}
a^{}_1 & \mapsto & * & * & * & \ldots & * & a^{}_{m} & a^{}_1 \\
a^{}_2 & \mapsto & . & . & . & \ldots & a^{}_{m} & a^{}_1 & . \\
\vdots &&&&& \diagup &\diagup &&\\
\vdots &&&& \diagup &\diagup &&&\\
a^{}_{m-1} & \mapsto & . & a^{}_{m} & a^{}_1 & \ldots &. & .& . \\
a^{}_{m} & \mapsto & a^{}_{m} & b^{}_1 & b^{}_1 & \ldots & b^{}_1 & b^{}_1 &
b^{}_1 \\  
b^{}_1 & \mapsto & * & * & * &  \ldots & * & a^{}_{m} &  b^{}_1 \\
\end{array}
\end{equation}
Here we have omitted ($.$) all letters that are not necessary and by $*$ we
denote the part that $\tilde{\theta}(a^{}_1)$ and $\tilde{\theta}(b^{}_1)$
share. We now check for \textit{pairwise coincidences}, i.e., for $i_1,
i_2 \in \ABC$ we check whether there is a $k \in \NN$ and a $j < \ell_{}^{k} =
(m+1)_{}^{k}$ such that $\sigma^{k}_{}(i_1)^{}_{j} =
\sigma^{k}_{}(i_2)^{}_{j}$. 

So we pick $i_1, i_2 \in \ABC=\{a^{}_1, \ldots, a^{}_m, b^{}_1\}$, $i_1 \neq
i_2$. Suppose $i_1 \neq a^{}_{m}$ (otherwise we interchange $i_1$ and
$i_2$). Then $i_2$ either equals $a^{}_{m}$ or at least $\tilde{\theta}(i_2)$
has an $a^{}_{m}$ (every $\tilde{\theta}(i)$, $i \in \ABC$ has one). In the
first case take $k=1$, otherwise $k=2$. Observe that there are $m$ successive
$b^{}_1$'s in $\tilde{\theta}(a^{}_{m})$. So, if we look at
$\tilde{\theta}_{}^{k}(i_1)$ and $\tilde{\theta}^{k}_{}(i_2)$, we get the
following: On the one hand, there are $m$ successive $b^{}_1$'s somewhere in
$\tilde{\theta}^{k}_{}(i_2)$, say at positions $j, \ldots, j+m-1$. On the
other hand, in $\tilde{\theta}_{}^{k}(i_1)$, there is at one of these positions
$j, \ldots, j+m-1$ either a $b^{}_1$, and we have a pairwise coincidence, or
an $a^{}_1$. Say there is an $a^{}_1$ at $\tilde{j}$ with $j \le \tilde{j} \le
j+m-1$. Then in $\tilde{\theta}_{}^{k+1}(i_1)$ and
$\tilde{\theta}_{}^{k+1}(i_2)$ we have pairwise coincidences at positions
$\tilde{j}\cdot\ell,\ldots,\tilde{j}\cdot\ell+m$ (the $*$'s
of~(\ref{eq:kol2m2_e})). 

From this pairwise coincidences we get a coincidence inductively: We start
with two letters $i_1, i_2$ and after $k_1\le 3$ substitutions we have a
pairwise coincidence, say at $j_1$. Now a third letter $i_3$ may have
something else at $\tilde{\theta}^{k_1}_{}(i_3)^{}_{j_1}$, but whatever it is,
in $\tilde{\theta}^{k_1+k_2}$ ($k_2 \le 3$) all three coincide somewhere at a
position $j_2$ with $j_1 \cdot \ell^{k_2}_{} \le j_2 < (j_1 +1)\cdot
\ell^{k_2}$. Since there are $\operatorname{card}(\ABC)=m+1$ letters, we get a
coincidence after at most $3\cdot m$ substitutions (i.e., there is a $j <
\ell^{3m}_{}$ such that all $\tilde{\theta}^{3m}_{}(i)^{}_{j}$ are the same
for all $i \in \ABC$).

The structure of $\theta$ is different. We have $\ABC =
\{a^{}_1,\ldots,a^{}_{m}, b^{}_2,\ldots b^{}_{n}\}$ and therefore \linebreak
$\operatorname{card} (\ABC)=m+n-1$. Let us first show an example, with $m=5$
and $n=3$:
\begin{equation}\label{eq:kol106}
\begin{array}{lcllllllll}
a^{}_1 & \mapsto & a^{}_1 & a^{}_2 & a^{}_3 & a^{}_4 & a^{}_5 & a^{}_1 &
a^{}_2 & a^{}_3 \\
a^{}_2 & \mapsto & a^{}_4 & a^{}_5 & a^{}_1 & a^{}_2 & a^{}_3 & a^{}_4 &
a^{}_5 & a^{}_1 \\
a^{}_3 & \mapsto & a^{}_2 & a^{}_3 & a^{}_4 & a^{}_5 & a^{}_1 & a^{}_2 &
a^{}_3 & a^{}_4 \\
a^{}_4 & \mapsto & a^{}_5 & a^{}_1 & b^{}_2 & b^{}_3 & a^{}_1 & b^{}_2 &
b^{}_3 & a^{}_1 \\
a^{}_5 & \mapsto & b^{}_2 & b^{}_3 & a^{}_1 & b^{}_2 & b^{}_3 & a^{}_1 &
b^{}_2 & b^{}_3 \\
b^{}_2 & \mapsto & a^{}_4 & a^{}_5 & a^{}_1 & a^{}_2 & a^{}_3 & a^{}_4 &
a^{}_5 & a^{}_1 \\
b^{}_3 & \mapsto & b^{}_2 & b^{}_3 & a^{}_1 & b^{}_2 & b^{}_3 & a^{}_1 &
b^{}_2 & b^{}_3 \\
\end{array}
\end{equation}
Since the positions of two consecutive $a^{}_1$'s in the sequence differ by at
most $m$, there is an $a^{}_1$ in every $\theta(i)$ with $i \in \ABC$ (note
that $\theta$ is a substitution of constant length $\ell=m+n$). Again we look
for pairwise coincidences, so choose $i_1, i_2 \in \ABC$. Then there is (at
least) one $a^{}_1$ in $\theta(i_1)$, say at position $j_1$, and (at least) one
in $\theta(i_2)$, say at position $j_2$. Since there can be more than one
$a^{}_1$ in either, we choose $j_1, j_2$ such that $|j_1-j_2|$ is minimal. We
further choose $i_1, i_2$ such that $j_1 < j_2$ (in the case $j_1 = j_2$,
e.g., $i_1=a^{}_1$ and $i_2=a^{}_{5}$ in the above example, we are already
done). If we look at $\theta(i_1)$ and $\theta(i_2)$, there are two cases each
(and therefore four cases, if we look at the combinations): Either
$\theta(i_1)^{}_{j_1+1} = a^{}_2, \ldots, \theta(i_1)^{}_{j_2} =
a^{}_{j_2+1-j_1}$ or $\theta(i_1)^{}_{j_1+1} = b^{}_2, \ldots,
\theta(i_1)^{}_{j_2} = b^{}_{j_2+1-j_1}$ and either $\theta(i_2)^{}_{j_1} =
a^{}_{m+1+j_1-j_2}, \ldots, \theta(i_2)^{}_{j_2-1} = a^{}_{m}$ or
$\theta(i_2)^{}_{j_1} = b^{}_{m+1+j_1-j_2}, \ldots, \theta(i_2)^{}_{j_2-1} =
b^{}_{m}$. This is all that can occur by the chosen minimality of $j_2-j_1>0$. 

Now we examine the case where $\theta(i_1)^{}_{j_1+1} = a^{}_2, \ldots,
\theta(i_1)^{}_{j_2} = a^{}_{j_2+1-j_1}$ and $\theta(i_2)^{}_{j_1} =
a^{}_{m+1+j_1-j_2}, \ldots, \theta(i_2)^{}_{j_2-1} = a^{}_{m}$. We want to
show that $\theta_{}^2(i_1)$ and $\theta_{}^2(i_2)$ have a pairwise
coincidence. Let us look at the $a^{}_1$'s in $\theta(a^{}_{i})$ only (we use
again the above example, but the reasons given apply for arbitrary $m$, $n$):
\begin{equation}\label{eq:ex-kol106}
{\renewcommand{\arraystretch}{1.7}
\begin{array}{lcllllllll}
a^{}_1 & \mapsto & \overset{1}{\bs{a}}^{}_1 & a^{}_2 & a^{}_3 & a^{}_4 &
a^{}_5 & \overset{2}{\bs{a}}^{}_1 & a^{}_2 & a^{}_3 \\
a^{}_2 & \mapsto & a^{}_4 & a^{}_5 & \overset{3}{\bs{a}}^{}_1 & a^{}_2 &
a^{}_3 & a^{}_4 & a^{}_5 & \overset{4}{\bs{a}}^{}_1 \\
a^{}_3 & \mapsto & a^{}_2 & a^{}_3 & a^{}_4 & a^{}_5 &
\overset{5}{\bs{a}}^{}_1 & a^{}_2 & a^{}_3 & a^{}_4 \\
a^{}_4 & \mapsto & a^{}_5 & \overset{6}{\bs{a}}^{}_1 & b^{}_2 & b^{}_3 &
\overset{7}{\bs{a}}^{}_1 & b^{}_2 & b^{}_3 & \overset{8}{\bs{a}}^{}_1
\\ 
a^{}_5 & \mapsto & b^{}_2 & b^{}_3 & \overset{9}{\bs{a}}^{}_1 & b^{}_2 &
b^{}_3 & \overset{10}{\bs{a}}^{}_1 & b^{}_2 & b^{}_3 \\
\end{array}}
\end{equation} 
First we number the $a^{}_1$'s with $1, \ldots, 2m$ (left to right in
$\theta(a^{}_{i})$ and top ($i=1$) to bottom ($i=m$)) and we will speak of the
$k$th $a^{}_1$ (with $1 \le k \le 2m$) according to that number. We observe
the following:
\begin{itemize}
\item Let $k<m$. If the $k$th $a^{}_1$ occurs at position $j \ge n$ in
  $\theta(a^{}_{i})$, then the $(k+1)$-st $a^{}_1$ occurs at position $j-n$ in
  $\theta(a^{}_{i+1})$. If the $k$th $a^{}_1$ occurs at position $j < n$,
  then the $(k+1)$-st $a^{}_1$ occurs at $j+m$ in the same $\theta(a^{}_{i})$.
\item Let $k>m+1$. If the $k$th $a^{}_1$ occurs at position $j \ge n$, then
  the $(k-1)$-st $a^{}_1$ occurs at $j-n$ in the same $\theta(a^{}_{i})$. If
  the $k$th $a^{}_1$ occurs at position $j<n$ in $\theta(a^{}_{i})$, then the
  $(k-1)$-st $a^{}_1$ occurs at position $j+m$ in
  $\theta(a^{}_{i-1})$.\footnote{Notice the contrary behaviour of the first
  two observations in going to a different or staying in the same
  $\theta(a^{}_{i})$ and the position of the corresponding $a^{}_1$. We call
  this the ``contrary line break property''.}
\item The second and the $(2m)$-th $a^{}_1$ occur at the same position $m$ in
  $\theta(a^{}_1)$, respectively $\theta(a^{}_{m})$. With the previous two
  observations we get: the $k$th and the $(2m+2-k)$-th $a^{}_1$ occur at the
  same position for $1 < k < m$.
\item The first and the  $(m+1)$-st $a^{}_1$ occur in $\theta(a^{}_{i})$ where
  there is at least one more $a^{}_1$. This is obvious for the first
  $a^{}_{i}$, for the $(m+1)$-st observe that if it occurs at position $j<m$,
  then there is also one at $j+n$, and if it occurs at position $j \ge m$,
  then there is also one at $j-m$. 
\end{itemize}
These observations are based on the facts that the length of the substitution
is $m+n$ and that the position of the $a^{}_1$'s in the sequence are separated
by $m$ or $n$ only. Now the fact that the $a^{}_{i}$ always occur in ascending
order (i.e., we have $a^{}_1a^{}_2a^{}_3\ldots$ and not
$a^{}_3a^{}_1a^{}_2\ldots$ or something else) together with the first two
observations essentially gives us an algorithm, which always yields a pairwise
coincidence in $\theta_{}^2(i_1)$ and $\theta_{}^2(i_2)$. Let us explain it in
our example~(\ref{eq:ex-kol106}): Suppose we have $i_1=a^{}_2$ and
$i_2=a^{}_3$. Then we have $j_1=2$ and $j_2=4$. The first step is always to
reduce $j_2$ by one, so we have $j_2'=3$. We have $j_2'\neq j_1$, but there
is a second $a^{}_1$ in $\theta(a^{}_{m})$ ($a^{}_{m}$ occurs at position
$j_2'$ in $i_2$!), so we can increment $j_1$ by $1$ and get $j_1'=3$. We
have $j_1'=j_2'$, and $\theta(\theta(a^{}_2)^{}_{j'_1})=\theta(a^{}_2)$ and
$\theta(\theta(a^{}_3)^{}_{j'_2})=\theta(a^{}_5)$ both have an $a^{}_1$ at
position $2$ (the third respectively the ninth $a^{}_1$). Therefore we get a
pairwise coincidence in $\theta_{}^2(a^{}_2)$ and $\theta_{}^2(a^{}_3)$.
This algorithm relies on the ``contrary line break property''.

The other three cases are mutatis mutandis the same, see the positions of the
$a^{}_1$'s in~(\ref{eq:kol106}). So, starting with any two $i_1, i_2 \in \ABC$
we get a pairwise coincidence in $\theta_{}^2(i_1)$ and
$\theta_{}^2(i_2)$. Inductively like before, we get a coincidence
after at most $2\cdot (m+n-2)$ substitutions. Therefore we have established
the following.

\begin{theorem}\label{thm:spectrum}
$\DS{\sigma}$ with $\sigma$ of~\textnormal{(\ref{eq:subs1})} has pure point
dynamical spectrum. Also, the dynamical system of the substitutions of
constant length as defined implicitly in~\textnormal{(\ref{eq:subs2})} and
$\theta, \tilde{\theta}$ of~\textnormal{(\ref{eq:subs3})} and
\textnormal{(\ref{eq:kol2m2})} have pure point dynamical spectrum.\qed 
\end{theorem}

\begin{varthm}[Proposition \ref{prop:spectrum}']
We have 
\begin{equation}
\DS{\sigma} \simeq \left\{ 
\begin{array}{ll} 
(\ZZ^{}_{m+n}\,\times\,\ZZ/2\ZZ,\tau) & \text{if }m+n\text{ is odd,}  \\
(\ZZ^{}_{m+n}, \tilde{\tau}) & \text{if }m+n\text{ is even},
\end{array} \right.
\end{equation}
where $\tau$ is the addition of $(1,1)$ and $\tilde{\tau}$ the addition of
$1$. \qed
\end{varthm}

\medskip
\noindent
\textsc{Remark}: Let $\varrho$ be a primitive substitution of constant length
$\ell$ and height $1$ over the alphabet $\ABC=\{1,\ldots,r\}$. Then we can
define the coincidence matrix $\bs{C}$, which is a quadratic \linebreak
$\frac12 r\cdot (r+1)\times \frac12r\cdot (r+1)$ matrix. The entries are
defined as follows (where $t\le s$ and $v \le u$):
\begin{equation*}
C_{(st)(uv)}=\left\{ \begin{array}{ll}
|\{j\mathbin| \varrho(s)^{}_{j} = u \; \wedge \; \varrho(t)^{}_{j} = u \}| &
\text{if } u=v \\
|\{j\mathbin| \varrho(s)^{}_{j} = u \; \wedge \;  \varrho(t)^{}_{j} = v \}| + 
|\{j\mathbin| \varrho(s)^{}_{j} = v \; \wedge \; \varrho(t)^{}_{j} = u \}|
& \text{if } u\neq v \\
\end{array} \right.
\end{equation*}
Note that the substitution matrix $\bs{M}$ is a submatrix of $\bs{C}$, since
$M^{}_{su} = C^{}_{(ss)(uu)}$. Also, $\bs{C}$ has row sums $\ell$. With this
definition, Lemma~\ref{lem:DekIII7} reads as follows.

\begin{prop}\cite[Proposition X.1]{Que87}\footnote{Note, however, that we use a
    definition of $\bs{C}$ different from~\cite{Que87}. The coincidence
    matrix there has dimension $r^2\times r^2$ and has the form (with the
    proper enumeration of the pairs)
\begin{equation*}
\left(\begin{array}{ccc}
M & 0 & 0 \\ R & P & Q \\ R & Q & P 
\end{array} \right)^{t}_{},
\end{equation*}
while the one defined above has the form
\begin{equation*}
\left(\begin{array}{cc}
M & 0 \\ R & P+Q  
\end{array} \right).
\end{equation*}
Here $M$, $P$, $Q$ are quadratic matrices and $M$ is the substitution
matrix. The Proposition is true for both matrices, the proof is analogous.} 
For $\DS{\varrho}$ are equivalent:
\begin{enumerate}
\item $\DS{\varrho}$ has pure point dynamical spectrum.
\item $\ell$ is a simple eigenvalue of the corresponding coincidence matrix
  $\bs{C}$.\qed
\end{enumerate}
\end{prop}
Obviously, $\ell$ is an eigenvalue of $\bs{C}$ ($\bs{C}/\ell$ is a stochastic
matrix with row sum $1$).

Now the above proof of Theorem~\ref{thm:spectrum} translates to the following
statements for $\bs{C}$:
\begin{itemize}
\item  For $\tilde{\theta}$, the third power of the coincidence matrix,
  $\bs{C}_{}^3$, has a column ($C^3_{(st)(a^{}_1a^{}_1)}$)  with nonzero
  entries only.  
\item For $\theta$, the square of the coincidence matrix,
  $\bs{C}_{}^2$, has a column ($C^2_{(st)(a^{}_1a^{}_1)}$) with nonzero
  entries only.
\end{itemize}

\begin{lem}\cite[Lemma X.3]{Que87}
Let $\bs{B}$ be a quadratic matrix with nonnegative integral entries and
row sums $\ell$. If $B_{ij} \ge 1$ for all $i$ and a fixed $j$, then $\ell$ is
a simple eigenvalue of $\bs{B}$.\qed
\end{lem}

This establishes the desired result that $\ell$ is a simple eigenvalue for the
coincidence matrix $\bs{C}$ of $\theta$ ($\tilde{\theta}$), since $\ell_{}^2$
($\ell_{}^3$) is a simple eigenvalue of $\bs{C}_{}^2$ ($\bs{C}_{}^3$).

We end this section with an example. We take $\tilde{\theta}$ for $m=2$,
$n=1$ and therefore the substitution
\begin{equation}\label{eq:kol42}
\begin{array}{lclll}
a^{}_1 & \mapsto & a^{}_1 & a^{}_2 & a^{}_1 \\
a^{}_2 & \mapsto & a^{}_2 & b^{}_1 & b^{}_1 \\
b^{}_1 & \mapsto & a^{}_1 & a^{}_2 & b^{}_1.
\end{array}
\end{equation}  
We get the coincidence matrix
\begin{equation*}
\bs{C} = \left( \begin{array}{cccccc}
2 & 1 & 0 & 0 & 0 & 0 \\
0 & 1 & 2 & 0 & 0 & 0 \\
1 & 1 & 1 & 0 & 0 & 0 \\
0 & 0 & 0 & 1 & 1 & 1 \\
1 & 1 & 0 & 0 & 1 & 0 \\
0 & 0 & 1 & 1 & 0 & 1   
\end{array} \right), \qquad
\bs{C}_{}^2 = \left( \begin{array}{cccccc}
4 & 3 & 2 & 0 & 0 & 0 \\
2 & 3 & 4 & 0 & 0 & 0 \\
3 & 3 & 3 & 0 & 0 & 0 \\
1 & 1 & 1 & 2 & 2 & 2 \\
3 & 3 & 2 & 0 & 1 & 0 \\
1 & 1 & 2 & 2 & 1 & 2   
\end{array} \right).
\end{equation*}
Here, already $\bs{C}_{}^2$ has columns with positive entries only. The
eigenvalues of $\bs{C}$ are \linebreak $\{0,0,1,1,2,3\}$.

\bigskip

\section{Model Sets and Diffraction}

A \textit{model set} $\varLambda(\Omega)$ (or \textit{cut-and-project set})
in \textit{physical space} $\RR^d$ is defined within the following general
cut-and-project scheme, see~\cite{Moo00,Baa02},
\begin{equation*} 
\renewcommand{\arraystretch}{1.5}
\begin{array}{ccccc}
& \pi & & \pi_{\textnormal{int}}^{} & \vspace*{-1.5ex} \\
\RR^{d} & \longleftarrow & \RR^{d}\times H & \longrightarrow & H \\
 & \mbox{\raisebox{-1.5ex}{\footnotesize
     \textnormal{1--1}}}\!\!\!\!\nwarrow\;\; & \cup &  
\;\;\nearrow\!\!\!\!\mbox{\raisebox{-1.5ex}{\footnotesize \textnormal{dense}}}
& \\
&  & \varGamma &  & 
\end{array}
\end{equation*}
where the \textit{internal space} $H$ is a locally compact Abelian group,
and $\varGamma\subset\RR^{d}\times H$ is a \textit{lattice}, i.e., a co-compact
discrete subgroup of $\RR^{d}\times H$. The projection
$\pi_{\textnormal{int}}^{}(\varGamma)$ is assumed to be dense in internal
space, and the projection $\pi$ into physical space has to be one-to-one on
$\varGamma$. The model set $\varLambda(\Omega)$ is 
\begin{equation*}
\varLambda(\Omega)\;=\;\left\{\pi(x)\mid x\in\varGamma,\,
\pi_{\textnormal{int}}^{}(x)\in\Omega \right\}\;\subset\; \RR^{d},
\end{equation*}
where the \textit{window} $\Omega\subset H$ is a relatively compact set with
nonempty interior.

Let $u$ be a bi-infinite sequence over $\ABC=\{1, \ldots, r\}$ and $\nu:\ABC
\to \CC, i \mapsto c^{}_{i}$ be a (bounded) function which assigns to every
letter a complex number (the \textit{scattering strength}). Then the
\textit{autocorrelation coefficients} $\eta(z)$ are given by
\begin{equation*}
\eta(z) = \lim_{N \to \infty} \frac{1}{2N+1} \sum_{n=-N}^{N}
\overline{\nu(u^{}_{n})} \cdot \nu(u^{}_{n+z})
\end{equation*}
provided the limits exist. We write $\delta^{}_{z}$ for the Dirac measure at
$z$, i.e., $\delta^{}_{z}(f)=f(z)$ for $f$ continuous. Then the
\textit{correlation measure} $\gamma$ of $u$ is given by
\begin{equation*}
\gamma = \sum_{z \in \ZZ} \eta(z) \, \delta^{}_{z},
\end{equation*}
and the \textit{diffraction spectrum}\footnote{So we think of $u$ as an atomic
chain, where there is an atom of type $u^{}_{n}$ at position $n$ with
scattering strength $\nu(u^{}_{n})$. We represent this atom as $\nu(u^{}_{n})
\cdot \delta^{}_{n}$ and therefore get a (countable) sum of weighted Dirac
measures with autocorrelation $\gamma$.} is given by the Fourier transform
$\hat{\gamma}$ of this measure. If $\hat{\gamma}$ is a sum of Dirac measures
only, i.e., $\hat{\gamma} = \sum_{k \in S} d^{}_{k} \cdot
\delta^{}_{k}$ with a countable set $S$ (for any choice of complex numbers
$(c^{}_{i})^{}_{i\in\ABC}$), then $u$ is \textit{pure point
diffractive}, i.e., the diffraction spectrum consists of
\textit{Bragg peaks} only. Also, the $d^{}_{k}$'s are the square of the
absolute value of the corresponding \textit{Fourier-Bohr coefficient} at $k$
and therefore non-negative (real) numbers. If there is no Dirac measure in
$\hat{\gamma}$, except one at position $0$, $\delta_0$, which is determined by
the density of the structure only, then the diffraction spectrum is
\textit{continuous}. 

For substitutive systems, the diffraction spectrum and
the spectrum of the corresponding dynamical system are closely related,
see~\cite{Dwo93, RW92, Rob96, LM01, LMS02}. 

\begin{prop}\label{prop:LM1}\cite[Corollary 1.]{LM01}
Let $\varrho$ be a primitive substitution of constant length $\ell$
with height $1$ over $\ABC=\{1,\ldots,r\}$, where $u$ is a fixed bi-infinite
word of $\varrho$. Define $U^{}_{i} = \{j\in\ZZ\mathbin| u^{}_{j} = i \}$ for
all $i \in \ABC$. We have $\ZZ = U^{}_1 \overset{.}{\cup} \ldots
\overset{.}{\cup} U^{}_{r}$, where $\overset{.}{\cup}$ denotes
disjoint union. Then the following are equivalent:
\begin{enumerate}
\item $\varrho$ admits a coincidence.
\item The $U^{}_{i}$'s are model sets for
\begin{equation*}
\renewcommand{\arraystretch}{1.5}
\begin{array}{ccccc}
& \pi & & \pi_{\textnormal{int}}^{} & \vspace*{-1.5ex} \\
\RR & \longleftarrow & \RR\times \ZZ^{}_{\ell} & \longrightarrow &
\ZZ^{}_{\ell} \\ 
\cup  & \mbox{\raisebox{-1.5ex}{\footnotesize
     \textnormal{1--1}}}\!\!\!\!\nwarrow\;\; & \cup &  
\;\;\nearrow\!\!\!\!\mbox{\raisebox{-1.5ex}{\footnotesize \textnormal{dense}}}
& \cup \\
\ZZ & \longleftarrow & \varGamma=\{(z,z)|z\in\ZZ\} & \longrightarrow & \ZZ
\end{array}
\end{equation*}
\item The sequence $u$ and the sets\footnote{By this we mean the
  special choice of $\nu$, where we set $c^{}_{i} = 1$ and $c^{}_{j} = 0$ for
  all $j\neq i$.} $U^{}_{i}$ are pure point diffractive.\qed
\end{enumerate}  
\end{prop}
Note that properties ``one-to-one'' and ``dense'' are obvious, the interesting
part is that there exist relatively compact windows (with respect to the
$\ell$-adic topology) with nonempty(!) interior. We will discuss this point
in the next section.

With this proposition, we know that $\theta$ and $\tilde{\theta}$ (as defined
implicitly in~(\ref{eq:subs3})) generate sequences which are pure point
diffractive. But we got every letter $i \in \ABC$ for the appropriate alphabet
for $\theta$, $\tilde{\theta}$ by building four-letter blocks in the
substitution rule~(\ref{eq:subs0}), e.g., in~(\ref{eq:kol42}) we have $a^{}_1
= 4444$, $a^{}_2 = 2222$ and $b^{}_1 = 4422$. Such a deterministic
substitution rule \linebreak
(i.e., Kol$(2m,2n)$ is \textit{local derivable} from the
sequence generated by $\theta$, respectively 
$\tilde{\theta}$) does not change the nature of the diffraction spectrum,
only the Fourier-Bohr coefficients. The diffraction spectrum
of Kol$(2m,2n)$ can be calculated from the one generated by
$\theta$, $\tilde{\theta}$ as follows: For Kol$(2m,2n)$ we only have
scattering strengths $c'_{2m}$ and $c'_{2n}$. If we therefore choose
$c^{}_{i}$ ($i \in \ABC$ with respect to $\theta$, $\tilde{\theta}$) according
to its four-letter-composition in $\{2m,2n\}$, then the diffraction spectrum
$\hat{\gamma}$ of $\theta$, $\tilde{\theta}$ is also a diffraction spectrum of
Kol$(2m,2n)$, where Kol$(2m,2n)$ is realized as an atom chain with atoms not on
$\ZZ$ but on $\frac14 \ZZ$ (we get the diffraction spectrum of Kol$(2m,2n)$
realized on $\ZZ$ by a simple rescaling with the factor $4$). For the
example~(\ref{eq:kol42}), this means that we choose
\begin{equation*}
\begin{split}
c^{}_{a^{}_1} = & \; c'_4 \cdot (1+e^{-\frac{2 \pi i}{4}}+ e^{-2 \frac{2 \pi
    i}{4}} + e^{-3 \frac{2 \pi i}{4}}) = 0 \\
c^{}_{a^{}_2} = & \; c'_2 \cdot (1+e^{-\frac{2 \pi i}{4}}+ e^{-2 \frac{2 \pi
    i}{4}} + e^{-3 \frac{2 \pi i}{4}}) = 0 \\
c^{}_{b^{}_1} = & \; c'_4 \cdot (1+e^{-\frac{2 \pi i}{4}}) + c'_2 \cdot (e^{-2
  \frac{2 \pi i}{4}} + e^{-3 \frac{2 \pi i}{4}}) = (1-i) (c'_4-c'_2). 
\end{split}
\end{equation*}
So in this case, the diffraction spectrum of Kol$(4,2)$ is given by the one of
$U^{}_{b^{}_1}$ only. All Kol$(2m,2n)$ are pure point diffractive.

\begin{lem}
For a sequence $u=\phi(v)$, where $v$ is a bi-infinite fixed
point of a primitive substitution and $\phi: \ABC^{}_{v} \to \ABC^{*}_{u}$ is a
morphism (where $u$ ($v$) is a sequence over $\ABC^{}_{u}$ ($\ABC^{}_{v}$)),
the following statements are equivalent: 
\begin{enumerate}
\item The dynamical system of $u$ has pure point dynamical spectrum.
\item $u$ has pure point diffraction spectrum.
\end{enumerate} 
\end{lem}

\begin{proof}
This is a (weak) conclusion of~\cite[Theorem 3.2]{LMS02}.
\end{proof}

We therefore obtain the following.

\begin{theorem}
All Kol$(2m,2n)$ have pure point diffraction and pure point dynamical
spectrum.\qed 
\end{theorem}

\medskip
\noindent
\textsc{Remarks}: The dynamical system of Kol$(2m,2n)$ is isomorphic to
\begin{equation}\label{eq:dynspec}
\begin{array}{ll}
(\ZZ^{}_{m+n}\,\times\,\ZZ/4\ZZ,\tau) \quad  & \text{if }m+n \equiv 1,3 \pod 4
\\ 
(\ZZ^{}_{m+n}\,\times\,\ZZ/2\ZZ,\tau) \qquad  & \text{if }m+n \equiv 2 \pod 4
\\ 
(\ZZ^{}_{m+n}, \tilde{\tau}) \qquad   & \text{if }m+n \equiv 0 \pod 4,
\end{array}
\end{equation} 
where $\tau$ is the addition of $(1,1)$ and $\tilde{\tau}$ the addition of
$1$. This can be seen by the fact that we get Kol$(2m,2n)$ from the
one generated by $\sigma$ of~(\ref{eq:subs1}) by the substitution $A = pp$,
$B=qq$, which corresponds just to doubling each letter in the latter one (see
the substitution matrices in~(\ref{eq:subs1}) and Footnote~\ref{foo:note}).

\medskip
\noindent
If we consider how to get from~(\ref{eq:subs2}), respectively~(\ref{eq:subs3}),
to Kol$(2m,2n)$ and compare this to Proposition~\ref{prop:LM1}, we get (compare
with~(\ref{eq:dynspec})): 
Kol$(2m,2n)$, respectively $U^{}_{2m}$, $U^{}_{2n}$ are model sets for 
\begin{equation*}
\renewcommand{\arraystretch}{1.5}
\begin{array}{ccccc}
& \pi & & \pi_{\textnormal{int}}^{} & \vspace*{-1.5ex} \\
\RR & \longleftarrow & \RR\times \ZZ^{}_{m+n} \times F & \longrightarrow &
\ZZ^{}_{m+n} \times F \\ 
& \mbox{\raisebox{-1.5ex}{\footnotesize
     \textnormal{1--1}}}\!\!\!\!\nwarrow\;\; & \cup &  
\;\;\nearrow\!\!\!\!\mbox{\raisebox{-1.5ex}{\footnotesize \textnormal{dense}}}
&  \\
& & \varGamma=\{(z,z,z \bmod{\operatorname{ord}(F)} )|z\in\ZZ\} & & 
\end{array}
\end{equation*}
where 
\begin{equation}\label{eq:cyclicgroup}
F \simeq \left\{ \begin{array}{ll} 
\ZZ/4\ZZ & \text{if } m+n \equiv 1,3 \pod 4 \\
\ZZ/2\ZZ & \text{if } m+n \equiv 2 \pod 4 \\ 
\{0\} & \text{if } m+n \equiv 0 \pod 4. 
\end{array}\right.
\end{equation}
(As before, $\ZZ^{}_{m+n}\simeq \ZZ^{}_{p^{}_1} \times \ldots \times
\ZZ^{}_{p^{}_{r}}$, where $p^{}_1, \ldots, p^{}_{r}$ are the distinct primes
dividing $m+n$.) The diffraction spectrum calculated from this cut-and-project
scheme is consistent with the previous one, since the Fourier-Bohr
coefficients which arise in each $\ZZ^{}_{m+n}$ separately are weighted by
factors $1,\, e^{-\frac{2 \pi i}{4}},\, e^{-2\frac{2 \pi i}{4}}$ or
$e^{-3\frac{2 \pi i}{4}}$ which depend on the element of the cyclic group
$\ZZ/4\ZZ$ (similar for the case $\ZZ/2\ZZ$), compare with~\cite{BKSZ90} ---
but this is just how we calculated the $c^{}_{i}$'s from $c'_{2m}$ and
$c'_{2n}$.

\medskip
\noindent
Let us show how one calculates the diffraction spectrum of $U^{}_{b^{}_1}$
of~(\ref{eq:kol42}) explicitly. This substitution can be written in recursive
equations for $U^{}_{a^{}_1}$, $U^{}_{a^{}_2}$ and $U^{}_{b^{}_1}$ by
observing at which position in which substitution a certain letter occurs
(e.g., $b^{}_1$ occurs in $\tilde{\theta}(a^{}_2)$ at positions $1$ and $2$
and in $\tilde{\theta}(b^{}_1)$ at position $2$)\footnote{Note that these
  recursive equations form an \textit{iterated function system} (IFS) in
  $3$-adic space, because multiplication by a factor of $3$ is a contraction
  in the $3$-adic topology. The closure of the windows in the $3$-adic
  internal space is therefore given by the unique compact solution of this IFS
  by (a generalized version of) Hutchinson's theorem~\cite[Section
  3.1(3)]{Hut81}. This method is well known for unimodular substitutions of
  Pisot-type, see~\cite{LGJJ93}, \cite{BS02} and the vast literature about
  Rauzy fractals. Similar results apply for all primitive substitutions of
  constant length $\ell$ in $\ell$-adic space.}:
\begin{equation*}
\begin{array}{lclclcl}
U^{}_{a^{}_1} & = & 3\, U^{}_{a^{}_1} & \cup & 3\, U^{}_{a^{}_1}+2 & \cup &
3\, U^{}_{b^{}_1} \\ 
U^{}_{a^{}_2} & = & 3\, U^{}_{a^{}_1}+1  & \cup & 3\, U^{}_{a^{}_2} & \cup &
3\, U^{}_{b^{}_1}+1 \\ 
U^{}_{b^{}_1} & = & 3\, U^{}_{a^{}_2}+1 & \cup & 3\, U^{}_{a^{}_2}+2 & \cup &
3\, U^{}_{b^{}_1}+2, 
\end{array}
\end{equation*} 
where $r \, U^{}_{i} + s = \{r\cdot z+s\mathbin| z \in U^{}_{i}\}$. Iterating
these equations and using $U^{}_{a^{}_1} \cup U^{}_{a^{}_2} \cup U^{}_{b^{}_1}
= \ZZ$, one gets:
\begin{equation*}
\begin{split}
U^{}_{b^{}_1} = &\; (9\, \ZZ + 5) \; \cup\; (27\, \ZZ + 17) \; \cup\; (27\,
\ZZ + 22) \; \cup \; (81\, \ZZ + 53) \\ 
& \; \cup\; (81\, \ZZ + 58) \; \cup \; (81\, \ZZ + 64) \; \cup \; (81\,
\ZZ + 65) \; \cup \; \ldots 
\end{split}
\end{equation*}
The Fourier transform of each lattice coset $\omega^{}_{r\,\ZZ+s} = \sum_{z
  \in \ZZ} \delta^{}_{r\cdot z+s}$ is easy to calculate:
\begin{equation*}
\widehat{\omega^{}_{r\, \ZZ+s}} = \frac1{r} \, e^{-2\pi i \, k \, s}\,
\omega^{}_{\ZZ/r}
\end{equation*}
Every Fourier-Bohr coefficients of $\widehat{U^{}_{b^{}_1}}$ are then given by
the sum of the Fourier-Bohr coefficients of the corresponding
$\widehat{\omega^{}_{r\, \ZZ+s}}$. The structure of $U^{}_{b^{}_1}$ (similar
for all $U^{}_{i}$ that occur for substitutions of constant length) as a union
of a countable but infinite set of (periodic) lattice cosets $r\cdot \ZZ+s$
gives rise to the name \textit{limit-periodic}, see~\cite{GK97}. 

\medskip
\noindent
The \textit{support} of the Bragg peaks of Kol$(2m,2n)$ is given by 
\begin{multline*}
\left\{ \left. \frac{k}{4 \cdot (m+n)^{s}}\,\right|\, k \in \ZZ,\; s\in
  \NN^{}_0 \right\} \\  = \left\{ \left.
\frac{k}{2^{\varepsilon}\cdot p^{s_1}_1 \cdot \ldots \cdot p^{s_{r}}_{r}}
  \,\right|\, k \in \ZZ,\; s_1,\ldots, s_{r} \in \NN^{}_0,\; \varepsilon \in \{
  0,\ldots, \log^{}_2(\operatorname{ord}(F))\} \right\}, 
\end{multline*}
where $p^{}_1,\ldots,p^{}_{r}$ are the distinct primes dividing $m+n$ and $F$
is the cyclic group of~(\ref{eq:cyclicgroup}). However, there need not be a
Bragg peak on every point of the support, e.g., Kol$(8,4)$ is equivalent to a
substitution $\theta$ of constant length $\ell = m+n = 6$, but the positions
of a letter $a^{}_{i}$, $b^{}_{i}$ are separated by multiples of
$2$: The support in this case is better described by $\{ \frac{k}{2^{s}}
  \mathbin| k \in \ZZ,\; s\in \NN^{}_0 \}$ than by $\{ \frac{k}{2^{s}\cdot
    3^{r}} \mathbin| k \in \ZZ,\; s,r \in \NN^{}_0 \}$.   

\bigskip

\section{Euclidean Models of \texorpdfstring{$\ell$}{l}-adic Internal Spaces}

So far, we have talked in an ``abstract'' way about the $\ell$-adic internal
space. Usually the discussion ends at this point, but we want to ``visualize''
this $\ell$-adic space. We hope that by doing this, we also gain some
intuition for such spaces and the meaning of $p$-adic internal spaces for
model sets. 

Recall that a $p$-adic integer can be written as a formal series $t=\sum_{i\ge
0} t^{}_{i} \cdot p^{i}_{}$ with integral coefficients $t^{}_{i}$ satisfying $0
\le t^{}_{i} \le p-1$ (\textit{Hensel expansion}). For the following, we
identify a $p$-adic integer $t$ with the sequence $(t^{}_{i})^{}_{i\ge 0}$ of
its coefficients. The set of all $p$-adic integers (a ring) is written as
$\ZZ^{}_{p}$, while the field of $p$-adic numbers is written as $\QQ^{}_{p}$
and can be seen as the set of all Laurent series $\sum_{i\ge N} t^{}_{i} \cdot
p^{i}_{}$ with $N \in \ZZ$. There is a $p$-adic \textit{valuation} $v^{}_{p}:
\QQ^{}_{p}\setminus\{0\} \to \ZZ$ defined by $v^{}_{p}(t) =
\min\{i\in\ZZ\mathbin|t^{}_{i} \neq 0 \}$, which gives rise to the $p$-adic
\textit{metric} with $|t|^{}_{p} = p^{- v^{}_{p}(t)}_{}$ and $|0|^{}_{p} = 
0$ (and $\QQ^{}_{p}$ and $\ZZ^{}_{p}$ are the completions of $\QQ$ and
$\ZZ$ with respect to the $p$-adic metric). So with respect to the $p$-adic
metric, two numbers in $\ZZ$ are close if their difference is divisible by a
high power of $p$. Note that this is a
\textit{non-Archimedean} absolute value (i.e., $|x+y|^{}_{p} \le
\max\{|x|^{}_{p}, |y|^{}_{p} \}$ for all $x,y \in \QQ^{}_{p}$) and we
therefore get some ``strange'' properties: all triangles are isosceles, every
point inside a ball $B_{r}(x) = \{ y \mathbin| |y-x|^{}_{p} < r\}$ is the
center of this ball, all balls are open and closed, etc., see~\cite{Gou97}. 
 
For Euclidean models (see~\cite[Section 1.2]{Rob00}) of $\ZZ^{}_{p}$ we only
need to know the formal series $t=\sum_{i\ge 0} t^{}_{i} \cdot p^{i}_{}$. We
can even show models for $\ZZ^{}_{\ell}$, where we do not make use
of~(\ref{eq:ladic}). For this, we use the addressing scheme known for fractals,
for example in the Sierpinsky gasket, see~\cite[Chapter IV.]{Bar88}: 
\begin{figure}[ht]
\centerline{\epsfxsize=0.65\textwidth\epsfbox{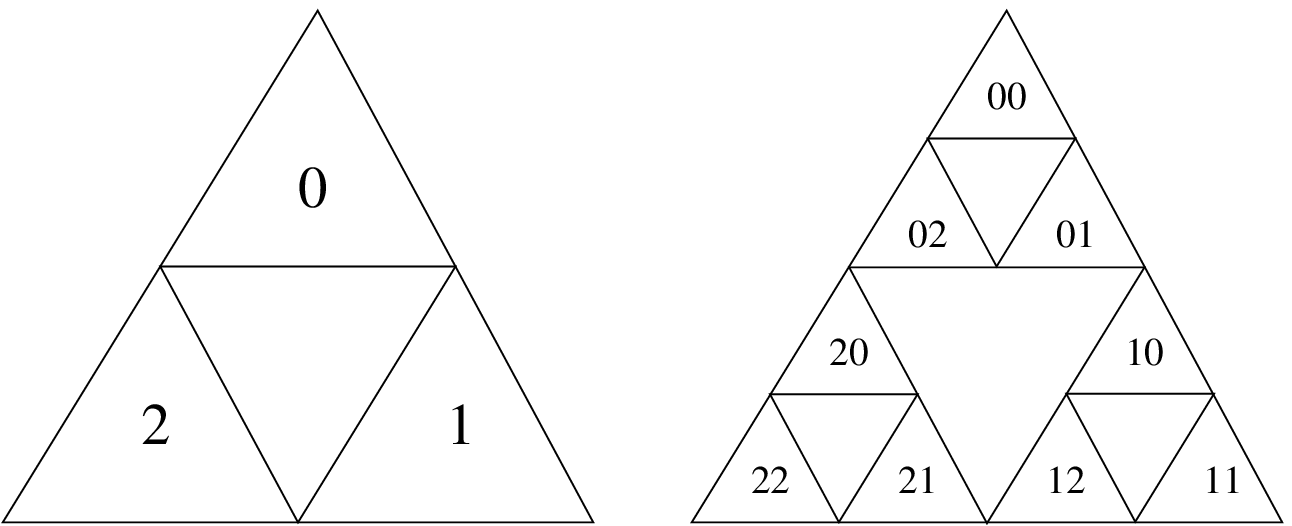}}
\end{figure}

\noindent
Now the interesting thing here is that each point in the Sierpinsky gasket has
a unique address (at least if we do not take the usual connected Sierpinsky
gasket but the totally disconnected version; this can be obtained by using a
contraction factor less than $\frac13$ in the IFS for the Sierpinsky
gasket). So each point in the Sierpinsky gasket corresponds to a sequence
$(t^{}_{i})^{}_{i\ge 0}$ with elements $0 \le t^{}_{i} \le 2$ --- this is just
the Hensel expansion of the $3$-adic integers. Similarly, the Cantor set is
such a geometric encoding of the $2$-adic integers. ``Reasonable'' geometric
representations of $\ZZ^{}_{\ell}$ in $\RR^{d}_{}$ are those, where the
sets $K^{}_{\{x_0\ldots x_r\}}=\{t \in \ZZ^{}_{\ell} \mathbin| t^{}_{0} =
x^{}_0, \ldots, t^{}_{r}= x^{}_{r}\}$  of points starting with the same
address are represented by objects of the same size for a fixed $r \in
\NN$. Therefore we get that in $d$-dimension, $\ZZ^{}_{\ell}$ with $d+1 \le
\ell \le (\text{kissing number in } \RR^{d}_{}) + 1$ can reasonably be
represented, if we do not make use of~(\ref{eq:ladic}). Note that we can
represent $\ZZ^{}_{3}$ either in $\RR^2_{}$ or $\RR$. 

This geometric representation surely fails for some $p$-adic (or $\ell$-adic)
properties (all triangles are isosceles, every point inside a ball is its
center, etc.), but some are also ``preserved'': points which are close in the
$p$-adic topology are also close in this geometric representation and the
representation as totally disconnected fractal corresponds to the totally
disconnected field $\QQ^{}_{p}$, $\ZZ^{}_{p}$ and its geometric models are both
compact sets. And balls in the $p$-adic topology correspond to scaled down
copies of the whole fractal.       

\begin{figure}[t]
\centerline{\epsfxsize=0.5\textwidth\epsfbox{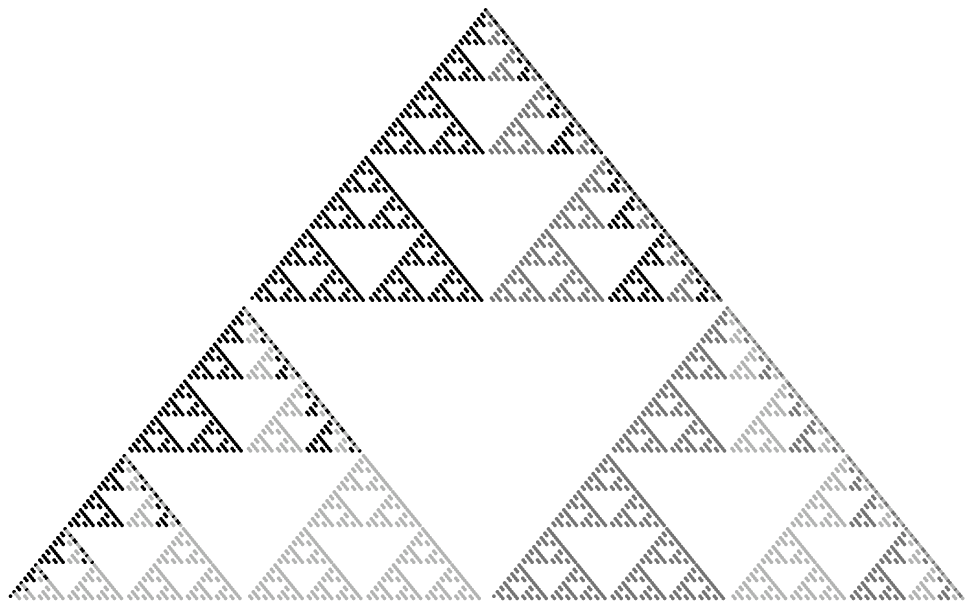}}\hspace{1ex}
\centerline{\epsfxsize=0.5\textwidth\epsfbox{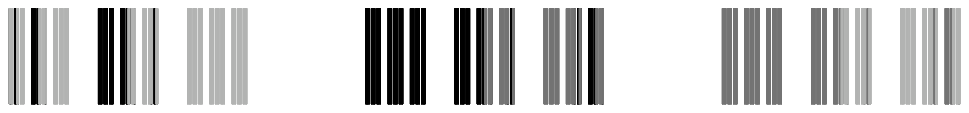}}
\caption{$3$-adic model for the internal space of~(\ref{eq:kol42}) in
  $\RR^2_{}$ (above) and $\RR$ (below, stretched for better
  representation). The colors correspond to $a^{}_1$ (black), $a^{}_2$ (dark
  gray) and $b^{}_1$ (light gray).\label{fig:kol42}} 
\end{figure}

We like to conclude this section with our example
from~(\ref{eq:kol42}). The $3$-adic geometric models are given in
Figure~\ref{fig:kol42}. Observe that, in the two-dimensional representation,
the parts (according to our above addressing scheme for the Sierpinsky gasket)
$K^{}_{\{02\}}$, $K^{}_{\{12\}}$ and $K^{}_{\{21\}}$ are colored by only one
color. This corresponds to the fact that at positions $9\, \ZZ+6$ are
$a^{}_1$'s, on $9\, \ZZ+7$ are $a^{}_2$'s and on $9\, \ZZ+5$ are $b^{}_1$'s
only in the bi-infinite sequence. So, large patches of the same color in the
geometric representations correspond to lattice cosets $\ell^{r}_{} \, \ZZ+s$
with small $r$. A similar addressing scheme can be used for the one-dimensional
representation (and in fact for all $\ell$-adic representations).  

\bigskip

\section*{Acknowledgments}

It is a pleasure to thank M.~Baake, R.V.~Moody and O.~Redner for discussions
and the German Research Council (DFG) for financial support.


\bigskip

\begin{thebibliography}{99}
\small

\bibitem{Baa02}
M.~Baake,
``A guide to mathematical quasicrystals'',
in \textit{Quasicrystals}, edited by J.-B.~Suck,  M.~Schreiber and
P.~H\"{a}ussler, Springer, Berlin, 2002, pp.\ 17--48;
\htmladdnormallink{math-ph/9901014}{http://arxiv.org/abs/math-ph/9901014}.  

\bibitem{BKSZ90}
M.~Baake, P.~Kramer, M.~Schlottmann and D.~Zeidler,
``Planar patterns with fivefold symmetry as sections of periodic structures in
4-space'', 
\textit{Int.\ J.\ Mod.\ Phys.} \textbf{B4} (1990), 2217--2268.

\bibitem{BS02}
M.~Baake and B.~Sing,
``Kolakoski-$(3,1)$ is a (deformed) model set'', 
preprint (2002); available at 
\htmladdnormallink{math.MG/020698}{http://arxiv.org/abs/math/0206098}.

\bibitem{Bar88}
M.F.~Barnsley,
``Fractals Everywhere'',
Academic, Boston, 1988.

\bibitem{Dek78}
F.M.~Dekking,
``The spectrum of dynamical systems arising from substitutions of constant
length'',
\textit{Z.\ Wahrscheinlichkeitstheorie verwandte Geb.} \textbf{41} (1978),
221--239.  

\bibitem{Dek97}
F.M.~Dekking,
``What is the long range order in the Kolakoski sequence?'',
in \textit{The Mathematics of Long-Range Aperiodic Order},
edited by R.V.~Moody, 
Kluwer, Dordrecht, 1997, pp.\ 115--125.

\bibitem{Dwo93}
S.~Dworkin,
``Spectral theory and X-ray diffraction'',
\textit{J.\ Math.\ Phys.}\ \textbf{34} (1993), 2965--2967.

\bibitem{BFMS02}
N.P.~Fogg, 
``Substitutions in Dynamics, Arithmetics and Combinatorics'',
\textit{Lecture Notes in Mathematics} \textbf{1784}, edited by
V.~Berth\'{e}, S.~Ferenczi, C.~Mauduit and A.~Siegel,
Springer, 2002; available at
\htmladdnormallink{http://iml.univ-mrs.fr/editions/preprint00/book/prebookdac.html}{http://iml.univ-mrs.fr/editions/preprint00/book/prebookdac.html}. 

\bibitem{GK97}
F.~G\"{a}hler and R.~Klitzing,
``The diffraction pattern of self-similar tilings'',
in \textit{The Mathematics of Long-Range Aperiodic Order}, 
edited by R.V.~Moody, 
Kluwer, Dordrecht, 1997, pp.\ 141--174.

\bibitem{Gou97}
F.Q.~Gouv\^ea,
``$p$-adic Numbers'',
2nd ed., Springer, Berlin, 1997. 

\bibitem{Hut81}
J.E.~Hutchinson,
``Fractals and self-similarity'',
\textit{Indiana Univ.\ Math.\ J.}\ \textbf{30} (1981), 713--747.

\bibitem{Kol65}
W.~Kolakoski,
``Self generating runs, Problem 5304'',
\textit{Am.\ Math.\ Monthly} \textbf{72} (1965), 674.

\bibitem{LM01}
J.-Y.~Lee and R.V.~Moody,
``Lattice substitution systems and model sets'',
\textit{Discrete Comput.\ Geom.}\ \textbf{25} (2001), 173--201;
\htmladdnormallink{math.MG/0002019}{http://arxiv.org/abs/math.MG/0002019}.

\bibitem{LMS02}
J.-Y.~Lee, R.V.~Moody and B.~Solomyak,
``Pure point dynamical and diffraction spectra'',
\textit{Annales Henri Poincar\'e} \textbf{3} (2003), 1003--1018;
\htmladdnormallink{mp\_arc/02-39}{http://www.ma.utexas.edu/mp_arc-bin/mpa?yn=02-39}.  

\bibitem{LGJJ93}
J.M.~Luck, C.~Godr\`{e}che, A.~Janner and T.~Janssen,
``The nature of the atomic surfaces of quasiperiodic self-similar
structures'',
\textit{J.\ Phys.\ A: Math.\ Gen.}\/ \textbf{26} (1993), 1951--1999.

\bibitem{Mar71}
J.C.~Martin,
``Substitution minimal flows'',
\textit{Am.\ J.\ Math.} \textbf{93} (1971), 503--526.

\bibitem{Moo00}
R.V.~Moody,
``Model sets: a survey'', 
in \textit{From Quasicrystals to More Complex Systems}, 
edited by F.~Axel, F.~D\'enoyer and J.P.~Gazeau,
EDP Sciences, Les Ulis, and
Springer, Berlin, 2000, pp.\ 145--166;
\htmladdnormallink{math.MG/0002020}{http://arxiv.org/abs/math.MG/0002020}.

\bibitem{Que87}
M.~Queff\'{e}lec,
``Substitution dynamical systems -- Spectral analysis'',
\textit{Lecture Notes in Mathenatics} \textbf{1294}, Springer, Berlin, 1987.

\bibitem{RW92}
C.~Radin and M.~Wolff,
``Space tilings and local isomorphism'',
\textit{Geom.\ Dedic.} \textbf{42} (1992), 355--360.

\bibitem{Rob96}
E.A.~Robinson, Jr.,
``The dynamical theory of tilings and quasicrystallography'',
in \textit{Ergodic Theory of $\ZZ^{d}$-Actions},
edited by M.~Pollicott and K.~Schmidt,
Cambridge U.\ P., Cambridge, 1996, pp.\ 451--473.

\bibitem{Rob00}
A.M.~Robert,
``A Course in $p$-adic Analysis'',
\textit{Graduate Texts in Mathematics} \textbf{198}, Springer, New York, 2000.

\bibitem{Diplom}
B.~Sing,
``Spektrale Eigenschaften der Kolakoski-Sequenzen'',
Diploma-Thesis, Universit\"{a}t T\"{u}\-bingen, 2002, available from the
\htmladdnormallink{author}{http://schubert.math-inf.uni-greifswald.de/sing/home.html}.  

\end{thebibliography}
\end{document}